# How many of the digits in a mean of 12.3456789012 are worth reporting?

R. S. Clymo*

School of Biological and Chemical Sciences, Queen Mary University of London, Mile End Road, London E1 4NS, UK

Email: clymo@rsjc.net; r.clymo@QMUL.ac.uk

**This version contains small amendments to that published as: Clymo R. S. (2019) *BMC Research Notes* 12 (148) https://doi.org/10.1186/s13104-019-4175-6**

## Abstract

Objective: A computer program tells me that a mean value is 12.3456789012, but how many of these digits are significant (the rest being random junk)? Should I report: 12.3?, 12.3456?, or even 10 (if only the first digit is significant)? There are several rules-of-thumb but, surprisingly (given that the problem is so common in science), none seem to be evidence-based.

Results: Here I show how the significance of a digit in a particular decade of a mean depends on the standard error of the mean (SEM). I define an index, $D_M$ that can be plotted in graphs. From these a simple evidence-based rule for the number of significant digits ('sigdigs') is distilled: the last sigdig in the mean is in the same decade as the first or second non-zero digit in the SEM. As example, for mean 34.63 ± SEM 25.62, with $n = 17$, the reported value should be 35 ± 26. Digits beyond these contain little or no useful information, and should not be reported lest they damage your credibility.

## Introduction

Numerous scientists – perhaps a majority – need to report mean values, yet many have little idea of how many digits carry useful meaning – are significant* ('sigdig's) – and at what point further digits are mere random junk. Thus a report that the mean of 17 values was 34.63 g with a standard error of the mean (SEM) of 25.62 g raises in a conspicuously permanent way a suspicion that none of the seven authors of the article were fully aware of what they were doing. But the frequency of a transition of a trapped and laser-cooled, lone ion of $^{88}Sr^{+}$ was reported [1] convincingly as 444 779 044 095 484.6 Hz, with an SEM of 1.5 Hz. It is a surprise that there seems to be no evidence to support the commonly used rules-of-thumb for this basic need. Here I derive simple evidence-based rules for restricting a mean value (and its SEM) to their sigdigs.

* ‘significant’ is used in this article with its general meaning: that this digit matters/is important/is not random. It is *not* a measure of statistical significance (conventionally $P <= 0.05$).

# Main text

**Table 1. Distribution of digits in a sample of 8000 values with mean 39.61500**

**A**: SEM 1.33

| Decade | Digit 0 | 1 | 2 | 3 | 4 | 5 | 6 | 7 | 8 | 9 | m$I_q$ |
|---|---|---|---|---|---|---|---|---|---|---|---|
| 10’s | ... | ... | ... | **4954** | 3046 | ... | ... | ... | ... | ... | 889 |
| 1’s | 1900 | 836 | 267 | 36 | 10 | 23 | 180 | 713 | 1727 | **2309** | 496 |
| 0.1’s | 785 | 785 | 764 | 841 | 807 | 751 | **827** | 851 | 813 | 776 | 193 |
| 0.01’s | 816 | **773** | 798 | 787 | 810 | 830 | 784 | 794 | 816 | 792 | 10 |
| 0.001’s | 849 | 782 | 809 | 818 | 766 | **790** | 820 | 792 | 775 | 799 | 13 |
| 0.0001’s | **809** | 789 | 781 | 817 | 815 | 782 | 831 | 803 | 771 | 802 | 11 |

By the 0.1’s the target digit is not the most frequent.

**B**: SEM 0.0133 (only 1/100 that in A above)

| Decade | Digit 0 | 1 | 2 | 3 | 4 | 5 | 6 | 7 | 8 | 9 | m$I_Q$ |
|---|---|---|---|---|---|---|---|---|---|---|---|
| 10’s | ... | ... | ... | **8000** | ... | ... | ... | ... | ... | ... | 1000 |
| 1’s | ... | ... | ... | ... | ... | ... | ... | ... | ... | **8000** | 1000 |
| 0.1’s | ... | ... | ... | ... | ... | 950 | **7050** | ... | ... | ... | 889 |
| 0.01’s | 1845 | **2330** | 1838 | 808 | 200 | 29 | 3 | 23 | 177 | 747 | 503 |
| 0.001’s | 823 | 802 | 828 | 783 | 790 | **770** | 831 | 786 | 787 | 800 | 12 |
| 0.0001’s | **818** | 766 | 822 | 812 | 778 | 796 | 814 | 823 | 793 | 778 | 12 |
| 0.00001’s | **788** | 834 | 788 | 817 | 788 | 839 | 841 | 807 | 741 | 757 | 10 |

Values drawn randomly from a Gaussian (‘normal’) population with mean 39.61500 and SEM as shown. The target digit in each decade is in **bold**; the most frequent digit in each row/decade is underlined. ‘...’ represents ‘0’. The sample of 8000 is an arbitrary choice that gives cell entries (in the lower rows) three digits. One measure of inequality along a row is $I_Q$ (the standardised sum of absolute differences from the row mean, range 0-1, see text), presented here multiplied by 1000 as m$I_Q$ .

## *Illustrative simulation*

To understand the trends, consider Table 1A which shows the frequency of digits in six decades (from the '10's to the '0.0001's) in 8000 random samples from a population of Gaussian ('normal') values with mean 39.61500 and SEM 1.33. In the 10's decade the frequency of '3's is a bit more than that of the '4's, reflecting the mean of 39. . .. The influence of the second digit ('9') is thus visible in the frequency of '4's in the '10's decade. The count (in **bold**) in target digit '3' is also the most frequent (underlined). This decade is clearly significant: one or more digits close to the target dominate the frequencies. The same is true of the '1's decade, though here there is a clear pattern of decline in frequency centred around the target '9'. In the '0.1's decade the target digit ('6') is only next to the most frequent digit ('7'), and pattern around '7' is not conspicuous.

We may measure inequality (non-uniformity) across the digits in a decade with an index, $I_Q$, based on the sum of absolute deviations from the mean in a row/decade, defined by the 'R' expression 'sum (abs (x – xbar)) / s', where x is a vector of the 10 counts for the individual digits, 0 to 9, xbar is the mean of the 'x' values, and 's = 2 * (sum (x) – mean (x)) = 1.8 * sum(x)' is a standardisation factor that brings $I_Q$ into the range 0 to 1. In Table 1 the $I_Q$ values are multiplied by 1000 as m$I_Q$. This $I_Q$ measure is linear and is a pure number, so values in different decades (rows) can be summed.

In Table 1A there are big reductions in $I_Q$ in the first three decades; thereafter values differ erratically governed by random frequencies of the digits. This pattern resembles an ice-hockey stick. As you move down the handle (rows/decades in Table 1A) the downward steps in the inequality measure are large. But when you reach the blade, differences in the measures between rows/decades become erratically smaller and larger, with no obvious further predictable change with additional rows/decades. At what decade may we suppose that little or no more useful information is present? This is tantamount to locating the junction between the hockey stick handle and blade. This is not a sharp angle, but a value of 20 mIq seems, from Table 1, to be suitable. A crude stopping-rule is thus to continue down the decades until m$I_Q$ is below 20, i.e. (Table 1A) to the same decade as the first or second digit in the SEM. This becomes Rule 1 in the Rules Box (later).

This rule uses the SEM to show where to stop: it makes no use whatever of the position of the decimal point. For example, the value 12.345 mm has 5 digits after the first non-'0', and 3 decimal places, while the same value in different units is 0.012345 m which also has 5 digits after the first non-'0' (i.e. ignoring preceding zeros) but 6, not 3, decimal places. Rules-of-thumb that specify a number of decimal places miss the point (literally as well as metaphorically) that precision is measured by SEM (and *n*).

Table 1B shows similar results for the same mean as in Table 1A, 39.61500, but SEM 100 times smaller. The same features are visible, and the same crude stopping-rule emerges. The '10's and '1's decades show only a single (the target) digit.; not until the '0.001's does $I_Q$ fall below 20 m$I_Q$.

The $I_Q$ calculation takes no notice of the *order* of the frequencies within a decade. Murray Hannah (personal communication) points out that at least one more decade may contain some residual conditional information. For example, in Table 1A, the 0.1's decade contains the 'run' of increasing or decreasing values 751, 827, 851, 813, 776, draped over the most frequent value: a faint echo of the strong patterns in earlier decades. But in Table 1B at the '0.001's decade (the first with m$I_Q$ < 20) there is no sign at all of a sequence. It seems that we need to add somewhere between 0 and 1 digits to the sigdig identified by the basic stopping rule (though this would require a fractional decade). At worst, the crude rule becomes *stop at the same decade as the second digit in the SEM*.

So far, so good. But two illustrative examples are far from sufficient to base a general rule on.

**A continuous index and trends for sigdigs.**

In Table 1A counts in the '0.1's decade show little regularity, but if we were to decrease the SEM gradually (details not shown) the totals for each digit in a decade become more and more unequal as frequency peaks emerge and grow from the hummocky sinking plain and, consequently, indicate that we may soon be able to justify another sigdig. The examples in Table 1 are indicative, but to understand the trends and to distil general rules, we need a sigdig index, $D_M$, for the mean that is continuous, and which can be plotted on a graph. For this purpose, because $I_Q$ is linear, we can simply add the $I_Q$ values for each decade (row) until we stop at the last decade with m$I_Q$ more than 20 ($I_Q$ more than 0.02). This value, $D_M = \Sigma I_Q$ , is then a plottable measure of sigdigs (Figures 1 and 2). In Figure 1, the large circles are for a stopping rule at 20 m$I_Q$, Putting the stopping rule at 10 m$I_Q$ (not shown) makes little difference. Sigdigs in the SEM, $D_{SEM}$ (Figure 2) are got in the same way as $D_M$.

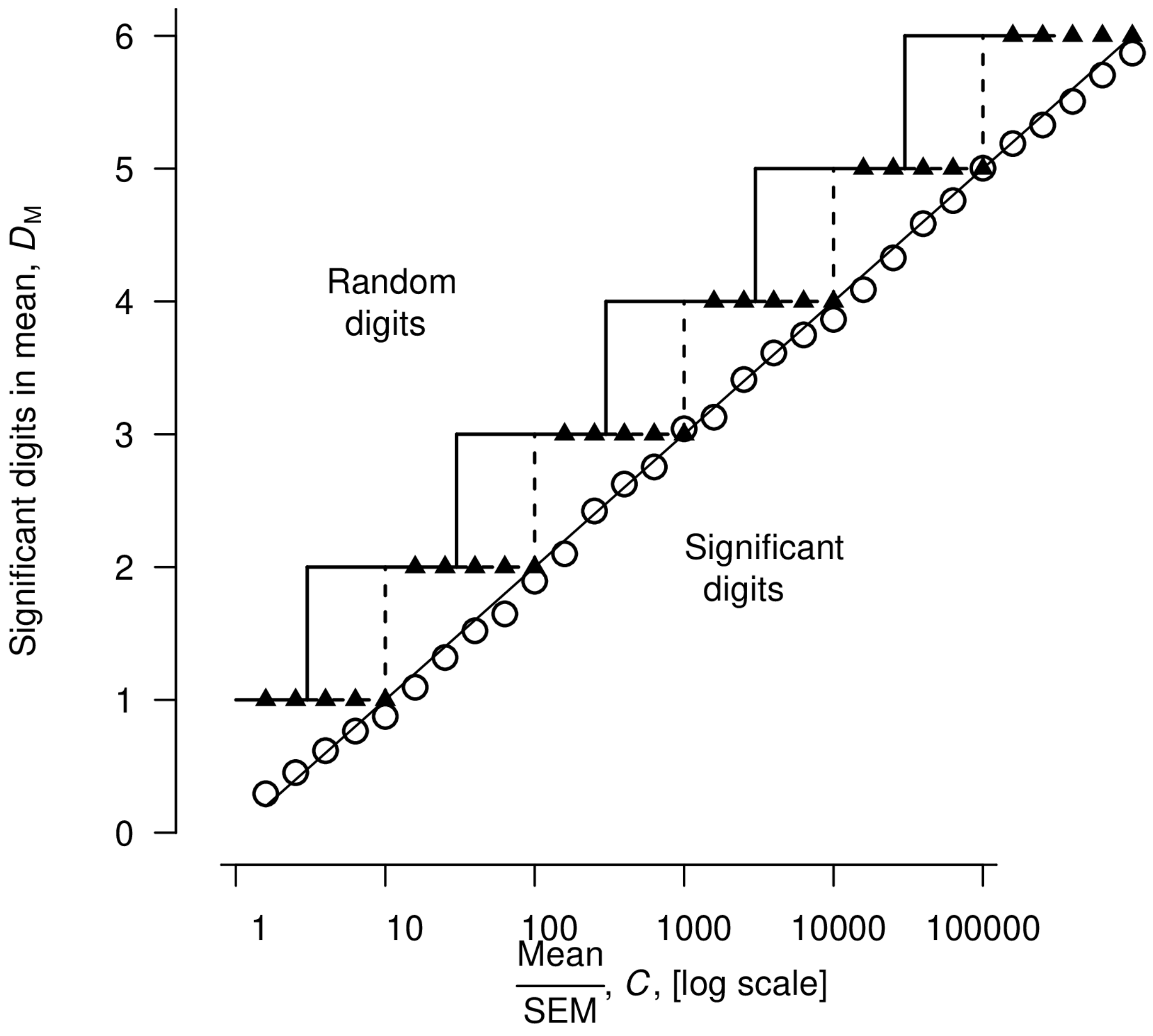

**Figure 1. Experimental dependence of sig-digs in a mean, $D_M$, on the $C$ = mean / SEM quotient.**

The small tringles are integer significant digits that are the sum of the decades reached so far. The unfilled circles are the index $D_M$ . They are close to the line $D_M = \log_{10}(C)$ line (with slope 1.0 and intercept zero). The area below and to the right of this line is the domain of significant digits in the mean; above and to the left the digits are random junk. The broken line staircase above the diagonal $D_M = \log_{10}(C)$ line shows the simple case for integer (1, 2, 3 and so on) sig-digs. Rule 1A (in the Rules Box, near the end) is for this broken line (see text). The unbroken line staircase shows the better but slightly more complex Rule 1B (listed in the Rules Box, near the end) that gives a more uniform distance between the staircase and the $D_M = \log_{10}(C)$ line.

## *Distilling rules*

The points in Figure 1 show how $D_M$ depends experimentally on $C$, the quotient of mean / SEM in experiments similar to those outlined in Table 1. The sloping line, $D_M = \log_{10} C$, is close to the circles, but is not fitted to them. If we take the ceiling of these values – equivalent to truncating and adding 1 – to get an integer value we get the broken line in Figure 1, superimposed on which is the direct integer sigdig (triangles). The overshoot into the random digits region is from 0 to 1 sigdig.

The possibility of Murray Hannah's contingent information may be accommodated by adding one extra decade to the dashed steps (Figure 1). It may be accommodated in another way: shift the steps about half a decade left using $\log_{10}(3) \approx 0.5$ (continuous line steps in Figure 1). The overshoot is more uniform at 0.5 to 1.5 digits, and this accommodates most if not all contingent information.

Rule 2 for $D_{SEM}$ is simpler but its origin is more complicated. Figure 2 shows, for a fixed mean and standard deviation (SD), how $D_{SEM}$ depends, in experiments similar to those in Table 1, on the number of items, $N_S$, in the calculation of an SEM. Points for two such experiments, with the same mean and different SDs are shown. Over a range of 100 the value of $D_{SEM}$ rises with a slope ≈ 1 on the log-linear scales shown: $D_{SEM} \approx \log_{10}(N_S) + c$, but eventually it falls over a cliff creating a sawtooth pattern. The cliff effect is at first very confusing. We know that the precision of the SD estimate must increase monotonically with increasing sample size. So too must the precision of the SEM. The reason for the cliffs is that, since SEM = SD / $\sqrt{N_S}$, it also decreases in magnitude. With every 100-fold increase in $N_S$ the SEM loses a leading significant decade, as a '1' in the leading decade shrinks to a '9' in the next decade. So while the precision increases, the number of significant digits decreases by one.

The overall slope of this saw-toothed progression (≈ 0.5) is half that of the teeth themselves reflecting the fact that the SEM depends on $\sqrt{N_S}$. The exact position of the sawtooth depends on the numerical value of the SEM, and to accommodate this the bounding line $D_{SEM} = \log_{10}(N_S) / 2 + 1$ is shown. The steps show Rule 2 (in the Rules Box). The offset for $N_S \leq 6$ accommodates the fact that at small $N_S$ the bounding line curves downwards, though this is not shown in detail in Figure 2. Reports of percentages have additional problems. The Rules Box below lists all these rules. Cole [2] considers the special case of risk (and other) ratios (strictly quotients).

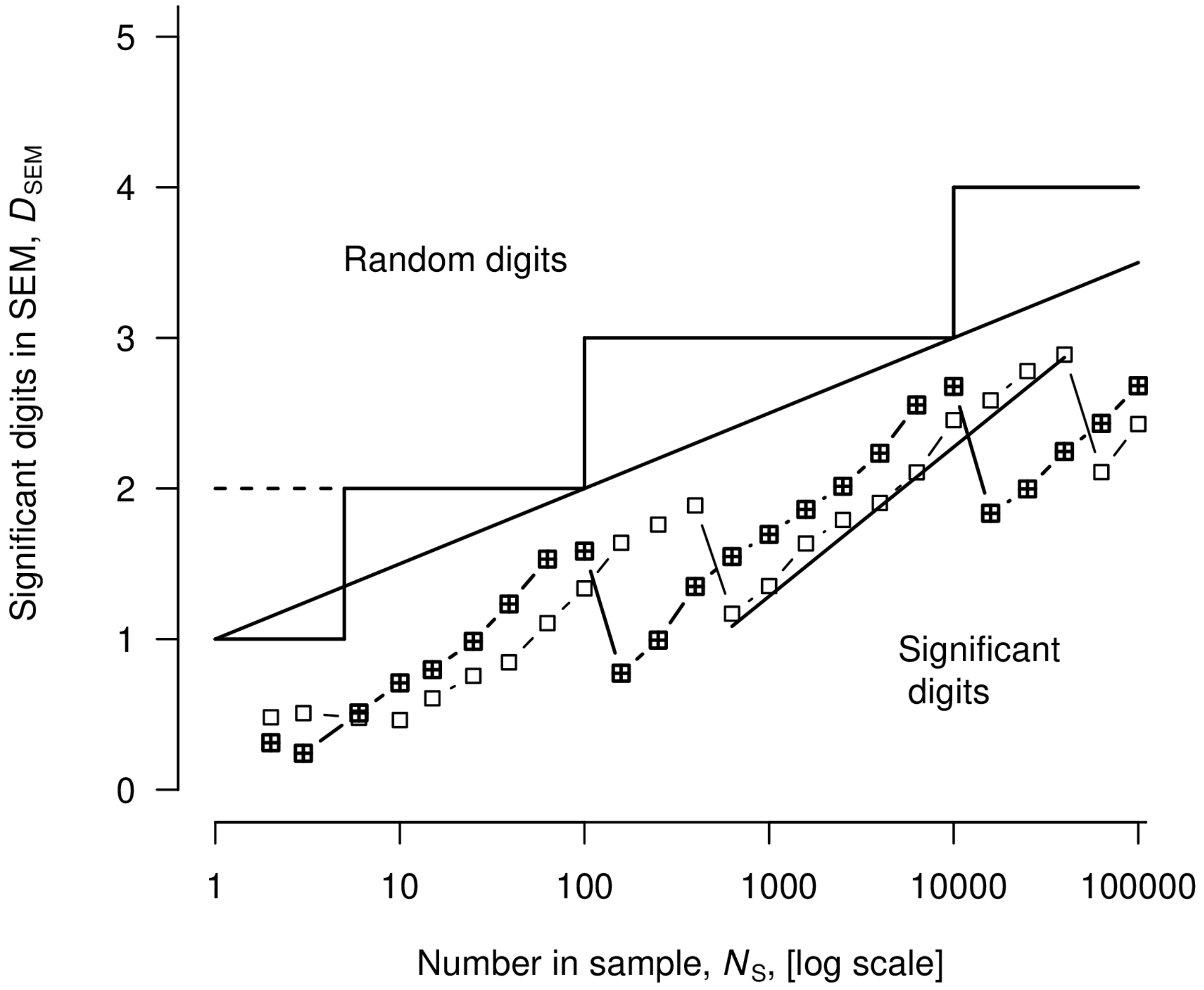

**Figure 2. Experimental dependence of sig-digs in the SEM, $D_{SEM}$, on the SEM sample size, $N_S$.**

For samples from a Gaussian population of $10^7$ with mean 39.5681 and the arbitrary standard deviation (SD) = 21.60 (empty squares) the points are close to a series of segments of lines $D_{SEM} = \log_{10}(N_S) + c$. The squares with a cross inside show a similar pattern for a SD half that used for the the empty squares. The line through the mean for one sawtooth has a slope of 1.0. The longer sloping line $y = \log_{10}(N_S) / 2 + 1$, with half the slope of the sawtooth lines, summarizes the upper bound of sawtooth lines and sets the boundary between significant and random digits. The staircase ending in a broken line, with a step every 100-fold increase in $N_S$ shows the simplest rule for significant digits in an SEM. The staircase with continuous lines and a short step at the bottom shows Rule 2 in the Rules Box, taking account of the difference in behaviour for very small $N_S$.

**Rules Box**

*Rule 1A: for significant digits ($D_M$) in the mean*:

The *last* significant digit in the mean is in the same decade as the *first* digit in the SEM; but, better is

*Rule 1B*

if the first significant digit in *C* = mean / SEM is '4' to '9' then, as in Rule 1A; but if *C* is '1' to '3' then the *last* significant digit in the mean is in the same decade as the *second* digit in the SEM.

*Rule 2: for significant digits ($D_{SEM}$) in the SEM itself:*

| *n* in sample | 2 to 6 | 7 to 100 | 101 to 10 000 | 10 001 to 1e6 | > 1e6 |
|---|---|---|---|---|---|
| Significant digits, $D_{SEM}$ | 1 | 2 | 3 | 4 | 5 |

*Rule 3: for counts as percentages*

For fewer than 100 observations then two digits in a percentage overstate the precision. For more than 100 (assuming counting statistics) *Rule 1* applies.

| *n* in sample* | 11 to 20 | 21 to 50 | 51 to 100 | 101 to 10 000 | 10 001 to 1e6 |
|---|---|---|---|---|---|
| Report % to the nearest / % | 5 | 2 | 1 | 0.1 | 0.01 |

* For 10 or fewer observations do not use %

Examples: 7 / 17 = 40 % (not 41.17. . . %); 6 / 17 = 35 %;

## *Special cases of zeros*

Suppose a raw mean of 0.0298699, has $D_M$ = 3 sigdigs under Rule 1A. The reported value should be 0.0300. The first two '0's locate the decade of the first sigdig; the final two '0's are significant, and their presence is sufficient to show that. They should not be omitted.

But suppose that the raw mean is 298699 with 3 sigdigs again, then the reported value should be 300000.The first two '0's are sigdigs, but the next 3 function only to show where the decimal point is. One way (there are others) to indicate such packing digits is by italics: 300*000*, or by expressing the value in exponent form: 3.00e5.

Finally, apply these rules to the example in the Introduction: mean = 34.63 , SEM = 25.62 , *n* = 17. This justifies SEM = 26, mean = 30 (Rule 1A) or 3*0* (the italic '*0*' is just a packing digit and its numerical value is not significant). Under Rule 1B, the mean would be 35 (mean / SEM = 1.35 so a further digit is significant)

## Limitation

This analysis deals with precision alone. Bias (and sometimes mistakes) may often have a bigger effect on a mean than does precision.

## Declarations

RSC declares that he has no competing interests, no funding, and is the sole author.

### *Acknowledgenents*

I thank Murray Hannah for pointing out possible contingent information beyond the $D_M$ limit. I salute those whose ignorance of when to stop goaded me to start this work.

## References.